\documentclass{appolb}
\usepackage{epsfig}

\begin{document}
\title{Theoretical understanding of pion production in nucleon-nucleon collisions - a status report
\thanks{Presented at the 7th International Workshop on Meson Production,
Properties and Interactions, MESON2002, Krakow, Poland.}%
}
\author{C. Hanhart
\address{Institut f\"ur Kernphysik, Forschungszentrum J\"ulich,
D-52425 J\"ulich}
}
\maketitle
\begin{abstract}
A status report is given for the current theoretical understanding of 
pion production close to threshold. In the first part of the talk 
predictions of a meson exchange model are compared to recent polarized data for
the reactions $pp \to pp\pi^0$ as well as $pp\to pn\pi^+$ revealing, that the
former reaction is badly described whereas the predictions for the latter turn
out to be consistent with the data. Recent progress in the application of
chiral perturbation theory allows to understand this difference.
\end{abstract}
\PACS{13.75.-n, 24.70.+s, 25.10.+s, 25.40.-h}
  
The highly accurate data for pion production in nucleon--nucleon collisions
close to the production threshold are a challenge for theoreticians.
When the first close to threshold data for the total cross section
of the reaction $pp \to pp\pi^0$ appeared in 1990, existing models fell short by
a factor of 5--10 \cite{KuR,MuS}. Many different mechanisms were proposed to cure this
discrepancy: heavy meson exchanges 
\cite{LuR}, (off-shell) pion rescattering 
\cite{HO,Han1}, 
excitations of baryon resonances \cite{Pena}, and  pion emission from
exchanged mesons \cite{CP2b}. The total cross sections
for the reactions $pp\to pn\pi^+$ and $pp\to d\pi^+$
on the other hand could always be described within a factor of 2 
--- the amplitude is dominated by the (on--shell) rescattering contribution \cite{KuR}.

\begin{figure}[t!]
\begin{center}
\epsfig{file=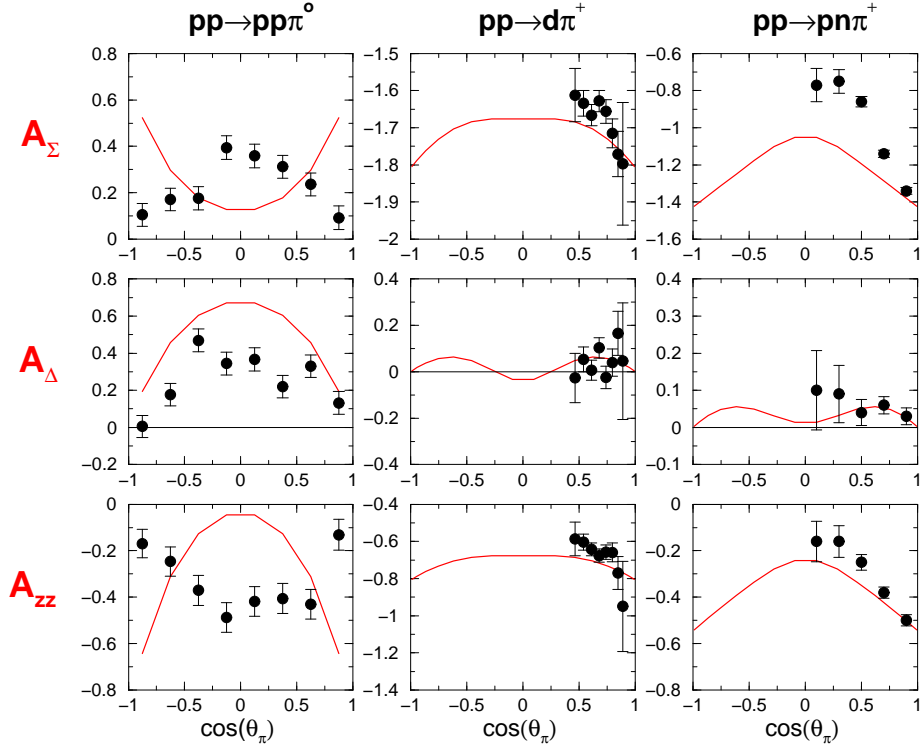, height=10cm, angle=0}
\caption{Comparison of the model predictions to the data taken
from Ref. \protect{\cite{polpppi0}} ($pp\to pp\pi^o$),
 Ref. \protect{\cite{poldpipl}} ($pp\to d\pi^+$) and
 Ref. \protect{\cite{polpnpipl}} ($pp\to pn\pi^+$).}
\label{erg}
\end{center}
\end{figure}

Recently the database on pion production was vastly enhanced due to
a large program at IUCF to measure double polarized observables for
the pion production reactions\cite{poldpipl,polpppi0,polpnpipl}. Unfortunately, until now there is
only one model published whose results can be compared to
those highly accurate data \cite{Han2}. This model includes 
the direct or one body terms, pion rescattering, where the 
interaction of the virtual pion with the second nucleon is 
taken from a microscopic model, and an additional diagram
that was introduced as an effective parameterization for the missing short range mechanisms.
The strength parameter of the latter is the only free parameter of
the model. It was adjusted to the total cross section
of $pp \to pp\pi^o$ close to the threshold.
After the publication of the data it turned out, that the model predictions
 are very successful
for the production channels involving charged pions whereas it badly fails for the differential
observables for the $\pi^o$ production (c.f. Fig. \ref{erg}). Thus, also 
here we find a striking difference between the production of charged and
neutral pions.

How can we understand this? A natural tool to use is that of effective
field theory, since the dynamics of the pions as  pseudo Goldstone
bosons is largely controlled by chiral symmetry. In the literature
there is a vast number of publications studying the $s$--wave of the
$pp \to pp\pi^o$ channel \cite{CP1,CPloop}. However, most of them employ the original
Weinberg counting scheme and thus ignore the large momentum scale inherent
to meson production reactions. The only work so far where this insight
was applied rigorously is Ref. \cite{chiralpwaves}. Here it was found
that the chiral expansion converges for pion $p$--waves and only tree 
level diagrams enter up to $N^2LO$. In the
case of $s$--waves, however, within this counting loops enter at $NLO$.
At leading order there is already a remarkable difference between the
production of $\pi^o$ and $\pi^+$: the leading order rescattering
vanishes in the former case whereas it is sizable in the latter,
driven by the so called Weinberg--Tomazawa term. For the leading loop
contributions we observe a similar pattern: the loops add up to zero
for the $pp \to pp\pi^o$ channel whereas there is a remainder for the
charged pions (note: at this order loops that contain a Delta
isobar add up to zero in both channels) \cite{mitnorbert}. At $N^2LO$ the number of
loop diagrams is quite large and a couple of low energy constants 
enter, that might be estimated from resonance saturation.

Thus we may conclude at this stage, that the reactions $pp \to pn\pi^+$
as well as $pp \to d\pi^+$ are well under control theoretically.  This
is not true for the reaction $pp \to pp\pi^o$. However, effective
field theory arguments can be used to understand this finding.


\begin{thebibliography}{10}
 
\bibitem{KuR} D. Koltun and A. Reitan, 
Phys. Rev. {\bf141}(1966)1413.

\bibitem{MuS} G. Miller and P. Sauer, 
Phys. Rev. C, {\bf 44}(1991)1725. 
 
\bibitem{LuR}
T.-S.H. Lee and D. Riska, Phys. Rev. Lett. {\bf 70},  2237  (1993);
C.J. Horowitz, H.O. Meyer, and D.K. Griegel,
Phys. Rev. C {\bf 49}, 1337 (1994).

\bibitem{HO}
E. Hern\'andez and E. Oset,  Phys. Lett. {\bf B350}, 158 (1995).

\bibitem{Han1}
C. Hanhart et al.,
Phys. Lett. {\bf B358},  21 (1995).

\bibitem{Pena}
M.T.~Pe\~na, D.O.~Riska, and A.~Stadler,
Phys. Rev. C {\bf 60} 045201 (1999).

\bibitem{CP2b}
U.~van Kolck, G.A.~Miller, and D.O.~Riska,
Phys.\ Lett.\ {\bf B388} 679 (1996).

\bibitem{poldpipl}
B.~von Przewoski {\it et al.},
Phys.\ Rev.\ C {\bf 58} (1998) 1897.
\bibitem{polpppi0}
H.~O.~Meyer {\it et al.},
Phys.\ Rev.\ C {\bf 63}, 064002 (2001).
\bibitem{polpnpipl}
W.~W.~Daehnick {\it et al.},
Phys.\ Rev.\ C {\bf 65} (2002) 024003

\bibitem{Han2}
C.~Hanhart et al.,
Phys.\ Rev.\ C {\bf 61} (2000) 064008


\bibitem{CP1}
T.D. Cohen, J.L. Friar, G.A. Miller, and U. van Kolck, 
Phys. Rev. C {53}, 2661 (1996); 
B.Y. Park et al., Phys. Rev. C {\bf 53}, 1519 (1996).
\bibitem{CPloop}
 E. Gedalin, A. Moalem, L. Razdolskaya, Phys. Rev. C {\bf 60}, 31 (1999),
V. Dmitrasinovic, K. Kubodera, F. Myhrer and  T. Sato,
Phys. Lett. {\bf B 465}, 43 (1999), 
S.~i.~Ando, T.~S.~Park and D.~P.~Min,
Phys.\ Lett.\ B {\bf 509} (2001) 253.

\bibitem{chiralpwaves}
C.~Hanhart, U.~van Kolck and G.~A.~Miller,
Phys.\ Rev.\ Lett.\  {\bf 85} (2000) 2905

\bibitem{mitnorbert}
C.~Hanhart and N.~Kaiser, in preparation.

\end{thebibliography}
\end{document}